\documentclass[epj]{svjour}
\usepackage{graphics}
\usepackage{amsmath,amssymb}
\usepackage{graphicx}
\usepackage[dvips]{epsfig}
\usepackage{subfigure}

\newcommand{\reaction}{\mbox{$pd \rightarrow{}^3 \textrm{He}\, \pi^+ \pi^-$}}

\newcommand{\reactionb}{\mbox{$dp \rightarrow{}^3 \textrm{He}\, \pi^+ \pi^-$}}
\begin{document}
\title{Isospin effects in the exclusive $\boldsymbol{dp\to{}^3}\textrm{He}\,\boldsymbol{\pi^+\pi^-}$ reaction}
\author{
M.~Mielke\inst{1}\thanks{Email: maltemielke@uni-muenster.de}\and
I.~Burmeister\inst{1}\and
D.~Chiladze\inst{2,3}\and
S.~Dymov\inst{2,4}\and
C.~Fritzsch\inst{1}\and
R.~Gebel\inst{2}\and
P.~Goslawski\inst{1}\and
M.~Hartmann\inst{2}\and
A.~Kacharava\inst{2}\and
A.~Khoukaz\inst{1}\and
P.~Kulessa\inst{5}\and
B.~Lorentz\inst{2}\and
T.~Mersmann\inst{1}\and
S.~Mikirtychiants\inst{2,6}\and
H.~Ohm\inst{2}\and
M.~Papenbrock\inst{1}\and
T.~Rausmann\inst{1}\and
V.~Serdyuk\inst{2}\and
H.~Str\"oher\inst{2}\and
A.~T\"aschner\inst{1}\and
Y.~Valdau\inst{2,7}\and
C.~Wilkin\inst{8}
}

\authorrunning{M.~Mielke \emph{et.al.}}
\titlerunning{Isospin effects in the exclusive $dp\to{}^3\textrm{He}\,\pi^+\pi^-$ reaction}

\institute{
%1
Institut f\"ur Kernphysik, Westf\"alische Wilhelms-Universit\"at M\"unster, D-48149 M\"unster, Germany \and
%2
Institut f\"ur Kernphysik and J\"ulich Centre for Hadron Physics, Forschungszentrum J\"ulich, D-52425 J\"ulich, Germany \and
%3
High Energy Physics Institute, Tbilisi State University, GE-0186 Tbilisi, Georgia \and
%4
Laboratory of Nuclear Problems, JINR, RU-141980 Dubna, Russia \and
%5
H.~Niewodniczanski Institute of Nuclear Physics PAN, PL-31342 Cracow, Poland \and
%6
High Energy Physics Department, Petersburg Nuclear Physics Institute, RU-188350 Gatchina, Russia \and
%7
Helmholtz-Institut f\"ur Strahlen- und Kernphysik, Universit\"at Bonn, D-53115 Bonn, Germany \and
%8
Physics and Astronomy Department, UCL, Gower Street, London WC1E 6BT, U.K.}
\date{Received: \today / Revised version:}
\abstract{The differential cross section for the exclusive \reactionb\
reaction has been measured with high resolution and large statistics
over a large fraction of the backward $^3$He hemisphere at the excess energy
265~MeV using the COSY-ANKE magnetic spectrometer. Though the well-known ABC
enhancement is observed in the $\pi^+\pi^-$ spectrum, the differences detected
between the $\pi^+{}^3$He and $\pi^-{}^3$He invariant-mass distributions show
that there must be some isospin-one $\pi\pi$ production even at relatively low
excess energies. The invariant-mass differences are modeled in terms of the
sequential decay $N^*(1440) \to \Delta(1232)\pi \to N\pi\pi$.
\PACS{{25.40.Qa}{$(p,\pi)$ reactions}
 \and {25.45.-z}{$^2$H-induced reactions}
 \and {25.40.Ve}{Other reactions above meson production thresholds (energies $>$ 400 MeV)}} % end of PACS codes
} %end of abstract
\maketitle

%
%%%%%%%%%%%%%%%%%%%%%%%%%%%%%%%%%%%%%%%%%%%%%%%%%%%%%%%%%%%%%%%%%%%%%%%%%%%%%%
%
\section{Introduction}
\label{intro}\setcounter{equation}{0}

Two-pion production in proton-deuteron collisions has a long history.
Abashian, Booth, and Crowe~\cite{ABA1963} measured the inclusive cross
sections for $pd \to{}^3\textrm{He}\,X^0$ and $pd \to{}^3\textrm{H}\,X^+$ at
a beam energy of $T_p=743$~MeV. This corresponds to an excess energy with
respect to the $\pi^+\pi^-$ threshold of
$Q=W-M_{^3\rm{He}}-2M_{\pi^+}=184$~MeV, where $W$ is the total energy in the
centre-of-mass system (CMS). In addition to the expected single-pion peaks, a
striking enhancement was seen in the $^3$He case at a missing mass of about
310~MeV/$c^2$ with a width $\approx 50$~MeV/$c^2$. This has since become
known as the ABC effect or enhancement. However, these parameters change with
the experimental conditions and a lot of evidence has since emerged to show
that the ABC is a kinematic effect, related to the presence of nucleons,
rather than an $s$-wave $\pi\pi$ resonance~\cite{AMS2012}.

The lack of a similar signal in the $^3$H case proves that the effect has to
be dominantly in the $\pi\pi$ isospin $I_{\pi\pi}=0$ channel. Apart from
phase space effects, one would then expect that the $\pi^+\pi^-$
component in the production of the ABC should be twice as strong as the
$\pi^0\pi^0$.

The original ABC data were obtained at a fixed laboratory angle of
$11.8^{\circ}$ and only covered forward production of the $^3$He with
respect to the proton beam direction in the CMS~\cite{ABA1963}. By using a
deuteron beam with an energy about twice as high, the acceptance was
increased significantly and allowed the reaction to be studied inclusively at
Saclay in both hemispheres~\cite{BAN1973}.

The first exclusive measurements of the \reaction\ reaction were made at much
lower energies and the data there look very different. The COSY-MOMO results
at $Q=76$~MeV, which did not differentiate between the charges on the two
pions, actually show a suppression of events at low $\pi^+\pi^-$ invariant
masses $M_{\pi^+\pi^-}$~\cite{BEL1999}. This was mistakenly taken as evidence
for the production of $p$-wave pion pairs. There is also no sign of an ABC
effect even closer to threshold in data obtained at $Q=27$~MeV at
CELSIUS~\cite{AND2000} or at 8 and 28~MeV at COSY-MOMO~\cite{JAH2006}.

Much more extensive information was provided by the CELSIUS-WASA
collaboration, where the production of both $\pi^+\pi^-$ and $\pi^0\pi^0$
pairs was measured at $Q=269$~MeV\footnote{All values of $Q$ are quoted with
respect to the $^3$He$\,\pi^+\pi^-$ threshold} in coincidence with the
detection of the $^3$He~\cite{BAS2006}. Only fast $^3$He were registered and
so the kinematics are similar to those of the first inclusive
experiment~\cite{ABA1963}. The large acceptance of WASA for the pions allowed
the measurement of a wide variety of angular distributions as well the
two-particle invariant masses.

The ABC effect was clearly seen in the WASA $\pi^0\pi^0$ spectrum but perhaps
somewhat less in that of $\pi^+\pi^-$. However, in the comparison of the two
final states there are significant effects at low $M_{\pi\pi}$ that arise
from the mass difference between the charged and the neutral pions.
\linebreak There are also some ambiguities in the fraction of $I_{\pi\pi}=1$
$\pi^+\pi^-$ pairs extracted from these data due to uncertainties in the
evaluation of the relative $\pi^0\pi^0/\pi^+\pi^-$ acceptance in WASA. However,
if one normalises the phase-space-corrected ratio to the isospin factor of
two in the low mass region, it appears that there could be a little
$I_{\pi\pi}=1$ production at very high $M_{\pi\pi}$. This is not in
contradiction to the $pd \to{}^3\textrm{H}\,X^+$ measurement~\cite{ABA1963}
since there are uncertainties in this case connected with possible background
contributions.

Both the $\pi^0\pi^0/\pi^+\pi^-$ and $pd \to{}^3\textrm{He}
\,X^0/{}^3\textrm{H}\,X^+$ cross section ratios are sensitive to the relative
$I_{\pi\pi}=1/I_{\pi\pi}=0$ production intensities. In contrast, the
difference in the $^3{}\textrm{He}\,\pi^+$ and $^3{}\textrm{He}\,\pi^-$
invariant mass distributions in the \reaction\ reaction depends upon the
interference of the amplitudes for producing $I_{\pi\pi}=1$ and
$I_{\pi\pi}=0$ $\pi^+\pi^-$ systems. There is therefore greater sensitivity
and, moreover, no uncertainty regarding the relative normalisations for
producing two different final states.

The CELSIUS-WASA collaboration evaluated experimental
$^3{}\textrm{He}\,\pi^+$ and $^3{}\textrm{He}\,\pi^-$
distributions~\cite{BAS2006} but did not draw firm conclusions from their
difference. We present here data taken at COSY-ANKE in inverse kinematics,
with a deuteron beam incident on a hydrogen target. This approach increases
significantly the overall acceptance and hence the statistics. The use of a
magnetic spectrometer also improves the resolution relative to WASA, though
this has to be balanced against the greater geometric acceptance at the WASA
facility.

The conditions for the \reactionb\ experiment are described in
sect.~\ref{experiment}, where the kinematics and the luminosity determination
are discussed. The event selection of sect.~\ref{selection} required the
detection of the $^3$He and at least one of the two pions. The reaction, or
the kinematics of the missing pion, was then identified with the help of the
missing-mass technique. By making judicious cuts on the data, it is believed
that fewer than 0.5\% of events were lost in the process but that the
background under the 81,500 good events was also below $\approx 1\%$.

The ANKE spectrometer has a very non-uniform acceptance and corrections for
this were estimated using a Monte Carlo model, where the input was taken from
two very different reaction models. These resulted in differences that were
typically less than 5\% and this was checked in sect.~\ref{acceptance} in a
model-independent multidimensional approach.

The results shown in sect.~\ref{results} are broadly in line with the
CELSIUS-WASA data~\cite{BAS2006} and are not incompatible with the inclusive
measurements at Saclay~\cite{BAN1973}. Apart from the ABC peak seen in the
$\pi^+\pi^-$ invariant mass distribution, both the $^3{}\textrm{He}\,\pi^+$
and $^3{}\textrm{He}\,\pi^-$ distributions show broad peaks that are
reminiscent of $\Delta(1232)$ excitation in nuclei. However, the high
resolution of the current experiment demonstrates quantitatively the shift in
the peak position between the $^3{}\textrm{He}\,\pi^+$ and
$^3{}\textrm{He}\,\pi^-$ data. This difference can be understood as arising
from the different coupling strengths of the $\pi^+$ and $\pi^-$ to the
$\Delta(1232)$ isobar, but this does not explain why this effect remains so
strong at low $M_{\pi^+\pi^-}$.

The mere existence of a $^3{}\textrm{He}\,\pi^+/^3{}\textrm{He}\,\pi^-$ mass
difference requires that there must be some $I_{\pi\pi}=1$ production. In our
conclusions of sect.~\ref{Conclusions}, we also discuss the relative
$I_{\pi\pi}=1/I_{\pi\pi}=0$ production intensities within the model used to
describe the $^3{}\textrm{He}\,\pi^+$ and $^3{}\textrm{He}\,\pi^-$
differences and also consider what information is available on this ratio
from the comparisons of $pd \rightarrow{}^3 \textrm{He}\, \pi^+
\pi^-/\pi^0\pi^0$ and $pd \rightarrow{}^3 \textrm{He}\,X^0/^3
\textrm{H}\,X^+$ data.
%
%%%%%%%%%%%%%%%%%%%%%%%%%%%%%%%%%%%%%%%%%%%%%%%%%%%%%%%%%%%%%%%%%%%%%%%%%%%%%%
%
\section{Experiment}
\label{experiment}\setcounter{equation}{0}

\begin{figure}[h]
\includegraphics[width=1.0\linewidth]{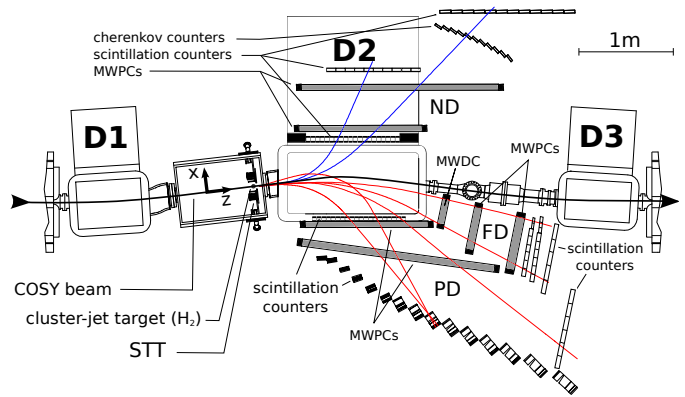}
\caption{The ANKE spectrometer setup showing the positions of the three
dipole magnets (D1,D2,D3), the hydrogen cluster-jet target, the Silicon
Tracking Telescope (STT), and the Positive (PD), Negative (ND), and Forward
(FD) detectors. } \label{ankesetup}
\end{figure}

The measurements were performed at the COoler SYnchrotron and storage ring
COSY~\cite{MAI1997} using the magnetic spectrometer ANKE~\cite{BAR2001} that
is located inside the ring. Two dipole magnets (D1, D3) deflected an
unpolarised deuteron beam onto a hydrogen cluster-jet target~\cite{KHO1999}
and back onto the nominal orbit, respectively (see Fig.~\ref{ankesetup}). The
third dipole magnet (D2) separated the charged reaction products from the
deuterons in the beam. The $^3\textrm{He}$ nuclei from the \reactionb\
reaction were registered in the forward detection (FD) system, the pions in the
two side detectors for positively (PD) and negatively (ND) charged particles.
Each of the detector systems is equipped with multiwire proportional chambers
and scintillator hodoscopes.

The principal aim of the experimental proposal was the determination of the
mass of the $\eta$ meson~\cite{GOS2012} and, for this purpose, the
measurements were carried out at 17 closely-spaced beam energies near the
$\eta$ production threshold. The experiment was divided into three
supercycles, each involving up to seven different machine settings. The
energy range covered is far above the threshold for two-pion production and
little variation is expected for this reaction over the narrow energy interval
$\Delta Q \sim 15$~MeV. The data presented in this work were therefore those
taken for background studies below the $\eta$ threshold, at a beam momentum of
$p_d=3.12$~GeV/$c$, where the highest statistics were collected\footnote{The
data actually used were taken from two supercycles with identical settings at
this energy.}.
This corresponds to $Q = 265$~MeV, which is effectively the same excess
energy as that used in the CELSIUS-WASA experiment~\cite{BAS2006}.

The kinematical range of the reaction measured at ANKE was restricted to CMS
$^3$He production angles from $143^{\circ}$ to $173^{\circ}$. In this region
there was practically full coverage of the \reactionb\ Dalitz plot.

The integrated luminosity that was necessary for the evaluation of
normalised cross sections was determined through the simultaneous measurement
of deuteron-proton elastic scattering, following the procedure described in
Ref.~\cite{MER2007}. After registering the scattered deuterons in the forward
detector, the reaction was isolated by identifying the proton with the
missing-mass peak. Calibration data from Refs.~\cite{DAL1968} were used in
the four-momentum-transfer region $0.08 < \left|t \right| <
0.26~(\textrm{GeV}/c)^2$.
The resulting normalisation has an overall systematic uncertainty of
$\pm6\%$.

%
%%%%%%%%%%%%%%%%%%%%%%%%%%%%%%%%%%%%%%%%%%%%%%%%%%%%%%%%%%%%%%%%%%%%%%%%%%%%%%
%
\section{Event selection}
\label{selection}\setcounter{equation}{0}

The events used in the analysis required a hit in the forward
detector associated with a hit in at least one of the two side
detectors. In cases where only one of the pions was
detected in coincidence with the $^3$He, the second pion was
identified through the missing mass in the reaction.

\begin{figure}[h!]
\includegraphics[width=1.0\linewidth]{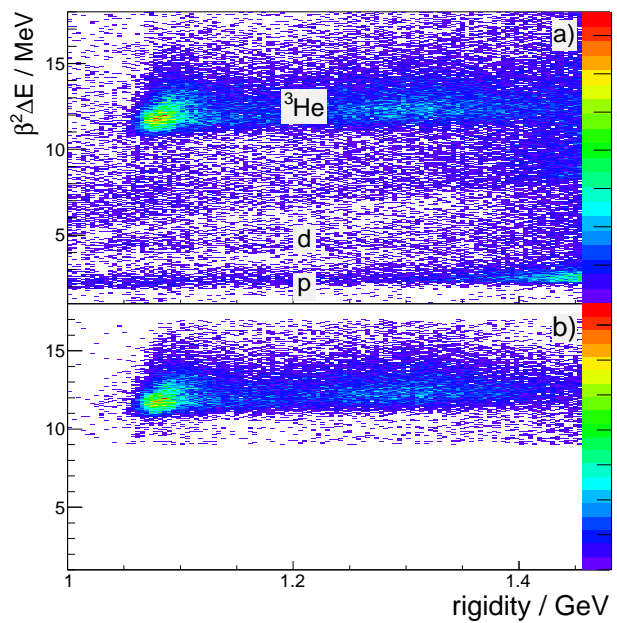}
\caption{(a) $\beta^2\Delta E$ versus rigidity before cuts. The
reconstructed particle rigidity is recorded on the abscissa. The ordinate
shows the energy deposited in the second scintillator wall of the forward
detector multiplied by the square of the particle velocity, on the assumption
that this is a $^3$He nucleus. The edge, which moves between 5.5 and 8.5~MeV,
is a consequence of the hardware trigger cut of about 16~MeV on $\Delta E$.
(b) The same spectrum after making cuts on the $^3$He band for each of the
three hodoscope layers.} \label{particleselection}
\end{figure}

Various steps were applied in the selection in order to identify the
\reactionb\ reaction. In the forward system a clear $^3\textrm{He}$ band is
apparent in the energy-loss-versus-momentum scatter plot shown in
Fig.~\ref{particleselection}a. As seen in Fig.~\ref{particleselection}b, a
large fraction of the background was suppressed by making cuts on
this band for each of the three hodoscope layers. Identical cuts were applied
to the Monte Carlo data used in the estimation of the acceptance corrections
(see section~\ref{acceptance}).

\begin{figure}[h]
\includegraphics[width=0.9\linewidth]{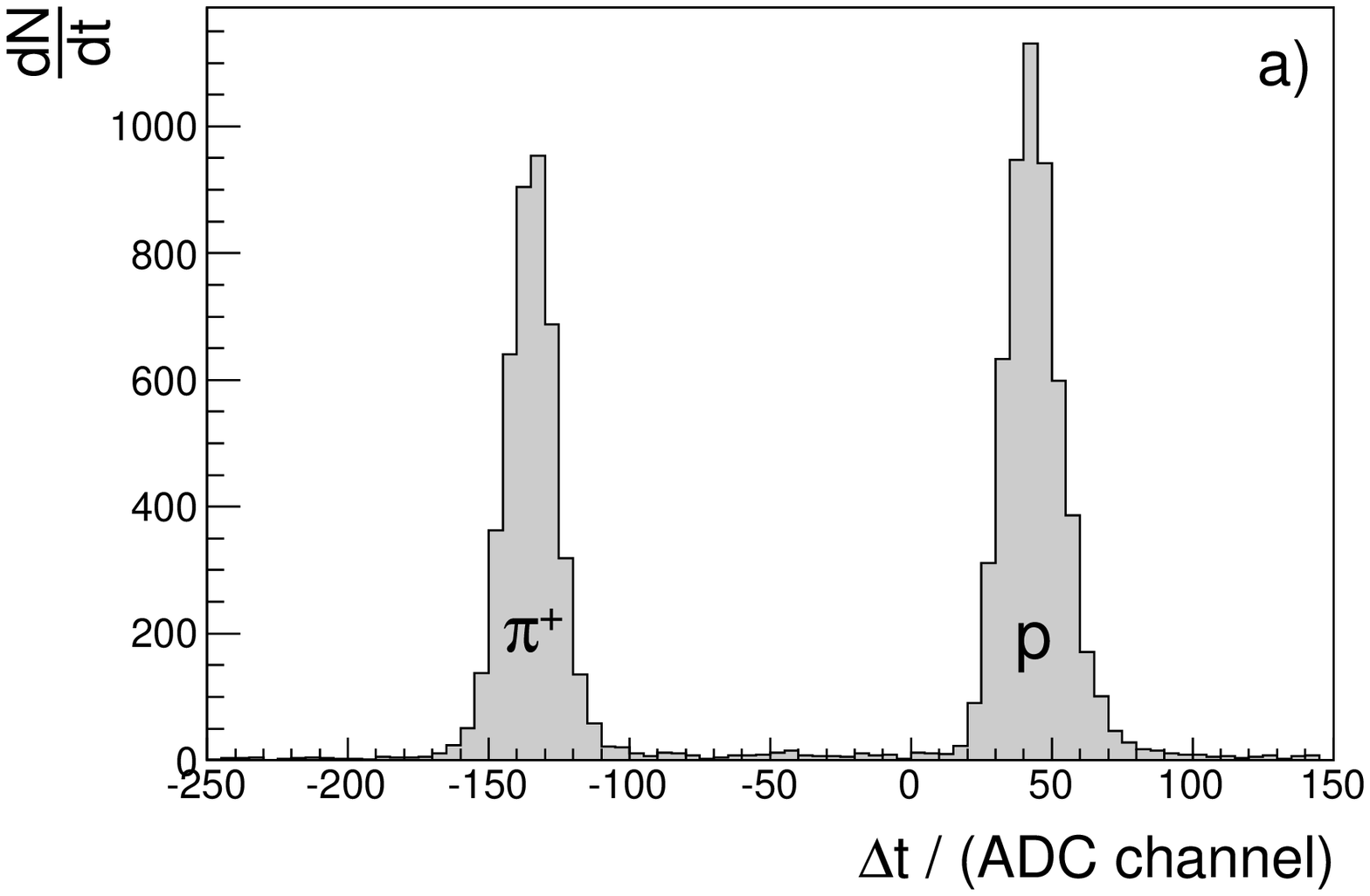}
\includegraphics[width=0.9\linewidth]{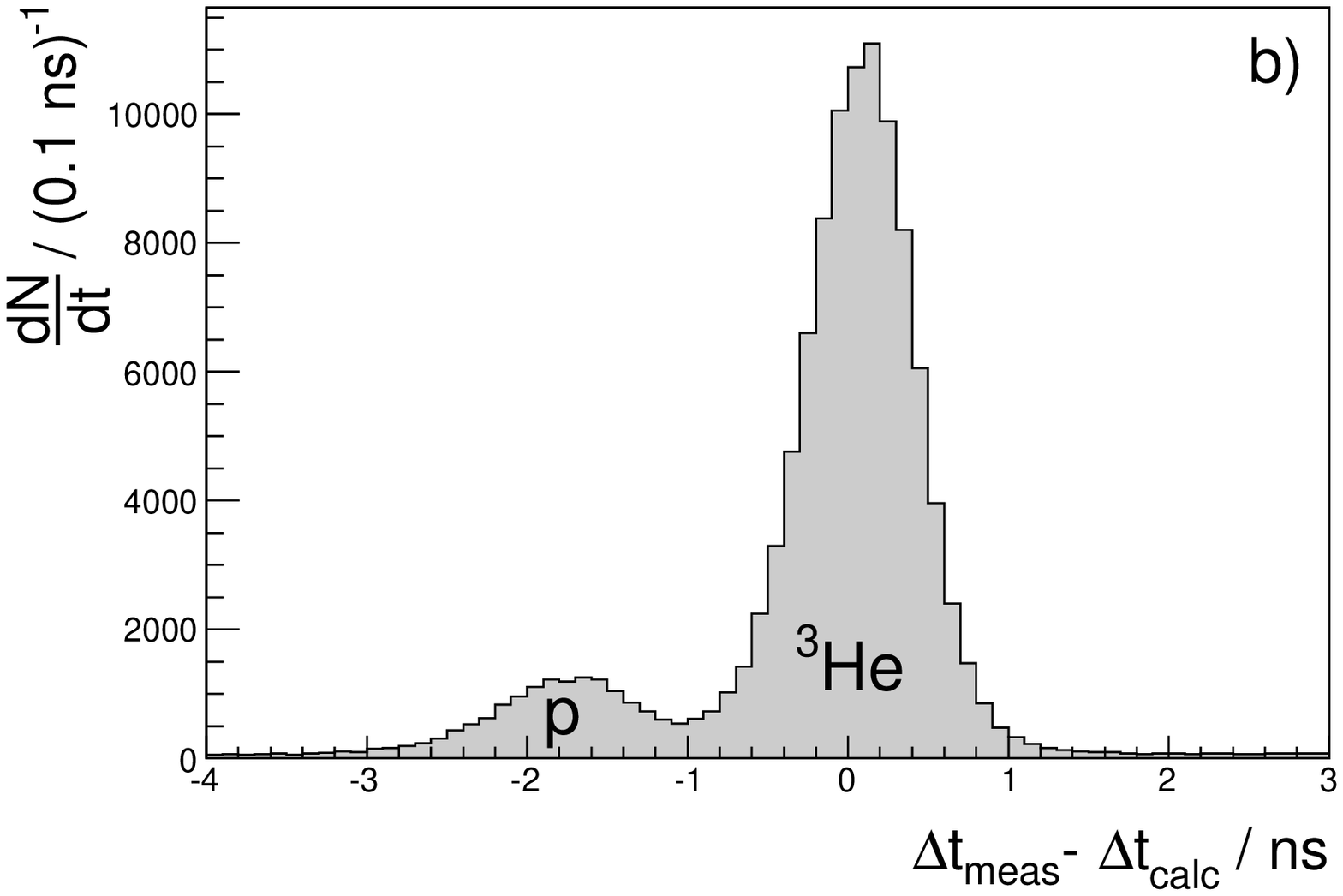}
\caption{(a) Time of flight difference between start and stop counters in the
positive detector. Pions are clearly separated from a background that
consists mainly of protons. (b) Difference between the times of flight
measured between the forward hodoscope and the start counter of a side
detector and that calculated from momentum information, on the assumption
that the particle is a $^3$He. The $^3$He peak, which is centered at the
origin, is well separated from a background that is composed mainly of
protons.} \label{particlecombiselection}
\end{figure}

The initial selection in the side detectors was made on the basis of the time
of flight (TOF) between start and stop counters. An example of this time
difference is shown in Fig.~\ref{particlecombiselection}a for a counter
combination involving the positive detector. A further cut on the TOF between
the start counters of the side detectors and the counters of the forward
detector was also imposed. To improve the resolution, the values of the times
recorded in the first two forward hodoscope layers were averaged and, as
shown in Fig.~\ref{particlecombiselection}b, this led to a good separation of
the $^3$He from the residual proton background. A $3\sigma$ limit was used
for all TOF selections.

\begin{figure}[h]
\includegraphics[width=1.0\linewidth]{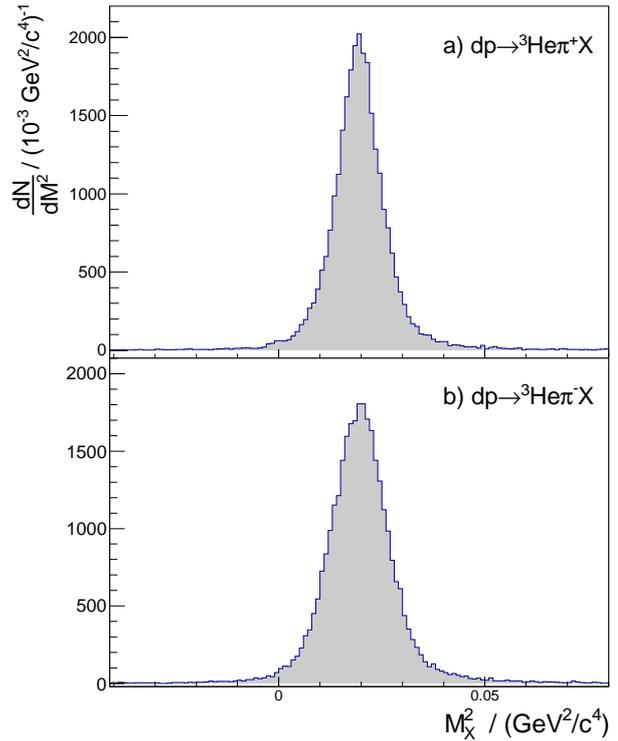}
\caption{Distributions in the squares of the missing masses for (a) $dp \to
{}^3\textrm{He}\,\pi^+X$ and (b) $dp \to {}^3\textrm{He}\,\pi^-X$ reactions.
Two-pion production can be clearly identified in both cases with minimal
background.}
\label{missingmass}
\end{figure}

The final selection of the \reactionb\ reaction was made using a missing-mass
cut of $\pm 0.03$~(GeV/$c^2)^2$ on the undetected pion. This is illustrated in
Figs.~\ref{missingmass}a and \ref{missingmass}b for unobserved $\pi^-$ and
$\pi^+$, respectively. When all three particles were detected a $4\sigma$ cut
on the invariant mass of $3.35$~GeV/$c^2$ was applied. Fewer than 0.5\% of the
events from the \reactionb\ reaction were lost by making these selections. The
total number of events, after making all cuts, was $\approx 81,500$, with a
background that is estimated to be below 1\%.

\vspace{-5mm}

%
%%%%%%%%%%%%%%%%%%%%%%%%%%%%%%%%%%%%%%%%%%%%%%%%%%%%%%%%%%%%%%%%%%%%%%%%%%%%%%
%
\section{Acceptance correction}
\label{acceptance}\setcounter{equation}{0}

In order to evaluate the \reactionb\ cross sections, a correction is required
to account for the limited acceptance of the ANKE detector. This has been
estimated through the use of Monte Carlo simulations. Due to the strong deviation of
the cross section from a simple phase space behaviour, combined with a
non-uniform detector acceptance, it is necessary to find a description which
reproduces adequately the main characteristics of the data.

There is no agreed model for the \reactionb\ reaction and the ansatz used
here is based loosely on the idea of the Roper resonance emitting a $p$-wave
pion and decaying into the $\Delta(1232)$ resonance, which also emits a
$p$-wave pion when it decays~\cite{Roper}. Neglecting recoil effects, the
simplest scalar that reflects the double $p$-wave transition is
$\vec{k}_1\cdot\vec{k}_2$, where the $\vec{k}_{i}$ are the pion momenta in
centre-of-mass frame. We are therefore led to consider the form
\begin{equation}
\sigma \propto \left|[M_{\pi}^{\,2} + B\vec{k}_1\cdot\vec{k}_2](3\Delta^{++}
+ \Delta^{0}) \right|^2,
\label{acccorrformula1}
\end{equation}
where the factors 3 and 1 result from the isospin couplings. The
$\Delta(1232)$ propagators are taken as~\cite{PIL1967}
\begin{equation}
\Delta = \frac{\sqrt{M_{\Delta}\Gamma/k_{\pi N}}}{M_{\pi N}^2-M_{\Delta}^2+iM_{\Delta}\Gamma},
\label{prop}
\end{equation}
where the width is parameterised by
\begin{equation}
\Gamma=b\frac{\gamma R^2k_{\pi N}^3}{1 + R^2k_{\pi N}^2},
\label{width}
\end{equation}
with $k_{\pi N}$ being the momentum in the $\pi N$ system.

Pion-nucleon data can be fitted with the values $M_{\Delta}=1.238$~GeV/$c^2$,
$\gamma =0.27$, and $R=6.3~c$/GeV~\cite{PED1978} but, in order to account for
a larger width of the isobar inside the $^3$He nucleus, a factor $b=1.35$ is
also introduced~\cite{WIL1973}. Our \reactionb\ data can then be reasonably
well described by Eq.~(\ref{acccorrformula1}) by choosing $B=0.2+0.3i$.

The primary purpose here is to provide a plausible ansatz in order to
estimate the necessary acceptance corrections. In order to study the model
dependence that this introduces, we consider also the more extreme
assumption;

\begin{equation}
\sigma \propto \left|3\Delta^{++}+ \Delta^{0}\right|^2P(M_{\pi^+\pi^-}).
\label{acccorrformula}
\end{equation}
The $p$-wave decays are omitted and a polynomial $P(M_{\pi^+\pi^-})$ included
to reproduce the shape of the $M_{\pi^+\pi^-}$ invariant mass spectrum;
\begin{equation}
P(x) = 1.00 - 6.05x + 12.52x^2 - 8.65x^3,
\end{equation}
where the masses are measured in GeV/$c^2$.

Despite the very marked differences between the forms of
Eqs.~(\ref{acccorrformula1}) and (\ref{acccorrformula}), the
acceptance-corrected distributions in the two models differ by at most 5\%.
Effects of a similar size are observed for modest variations of the 3:1
factor for the $\Delta^{++}$ and $\Delta^{0}$, as well as the parameters
$M_{\Delta}$ and $b$ in Eqs.~(\ref{prop}) and (\ref{width}).

In order to investigate these corrections further, estimates were also made
using a multidimensional approach in terms of the relevant degrees of
freedom, as described in Ref.~\cite{BAL2001}. For a three-particle final
state produced in an unpolarised reaction, there are four independent
observables. These were chosen to be the invariant masses of the
$^3\textrm{He}\,\pi^-$ and the $\pi^+ \pi^-$ systems, the $^3$He production
angle, and the polar angle of the normal to the ejectile plane, though the
latter had minimal influence on the results. The invariant masses were each
divided into ten bins and the $^3$He angle into six.

In certain kinematical regions, where there is very low and strongly
fluctuating acceptance, the multidimensional matrix method was not reliable.
Thus, when comparing the acceptance corrections with those based on
Eq.~(\ref{acccorrformula1}), the region where $3.03~\textrm{GeV}/c^2 <
M_{^3\rm{He}\,\pi^-} < 3.09~$GeV/$c^2$ and $M_{\pi^+\pi^-} > 0.43$~GeV/$c^2$
was excluded. Away from this region, both methods are in very good agreement,
which suggests that Eq.~(\ref{acccorrformula1}) could provide a useful basis
for evaluating one-dimensional acceptance corrections for the complete
invariant mass region.

The systematic uncertainties associated with the choice of model used for the
acceptance corrections were evaluated individually for each data point in the
respective invariant mass spectrum. This was done by comparing the correction
factors resulting from Eq.~(\ref{acccorrformula1}) with those obtained with
various model variations that are essentially also in agreement with the
data. These are Eq.~(\ref{acccorrformula}) and the modifications of the
Breit-Wigner functions that are mentioned above. For each correction factor
the maximum relative deviation was taken as a measure of the systematic
uncertainty.

%
%%%%%%%%%%%%%%%%%%%%%%%%%%%%%%%%%%%%%%%%%%%%%%%%%%%%%%%%%%%%%%%%%%%%%%%%%%%%%%
%
\section{Results}
\label{results} \setcounter{equation}{0}

After correcting for inefficiencies in the data acquisition system and the
MWPCs, the value of the \reactionb\ differential cross section has been
extracted at the excess energy $Q = 265$~MeV. Averaged over the $^3\rm{He}$
production angle interval $143^{\circ} < \vartheta^{CMS}_{^3\rm{He}} <
173^{\circ}$, this gave $\langle d\sigma / d\Omega^{CMS}\rangle = 480\pm 3
\pm 35$~nb/sr, where the first error is statistical and the second
systematic. The latter arises mainly from the 6\% uncertainty in the
luminosity, though there is also some contribution coming from the ambiguity
in the evaluation of the acceptance. The CELSIUS-WASA collaboration quoted a
value of $\langle d\sigma/d\Omega^{CMS} \rangle= 660\pm60$~nb/sr when
averaged over $141^{\circ} < \vartheta^{CMS}_{^3\rm{He}} <
180^{\circ}$~\cite{BAS2006}. However, their cross section normalisation was
derived from a comparison with $dp\to{}^3\textrm{He}\,\pi^0$ data, which
introduces a $\approx 20\%$ systematic uncertainty~\cite{BAS2014}.

The comparison with the Saclay inclusive data~\cite{BAN1973} is less
straightforward because only $180^{\circ}$ results are available in our
energy region and the quoted $560\pm70$~nb/sr corresponds to only the ABC
peak over the hand-drawn background. However, after taking into account the
$\approx 60\%$ $\pi^+\pi^-$ branching ratio, no obvious discrepancy is
evident with the present data.

\begin{figure}[h]
    \includegraphics[width=1.0\linewidth]{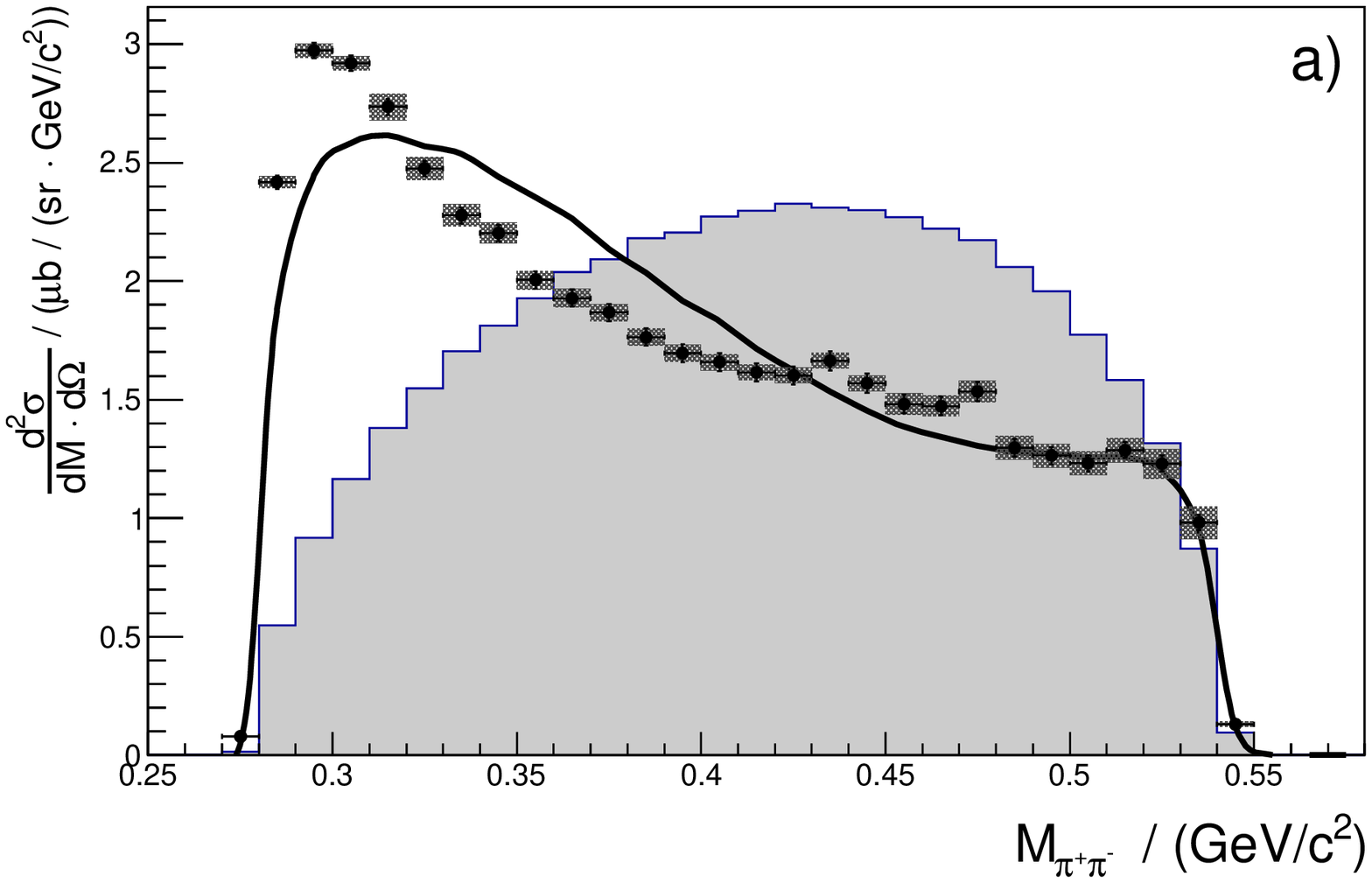}
	\includegraphics[width=1.0\linewidth]{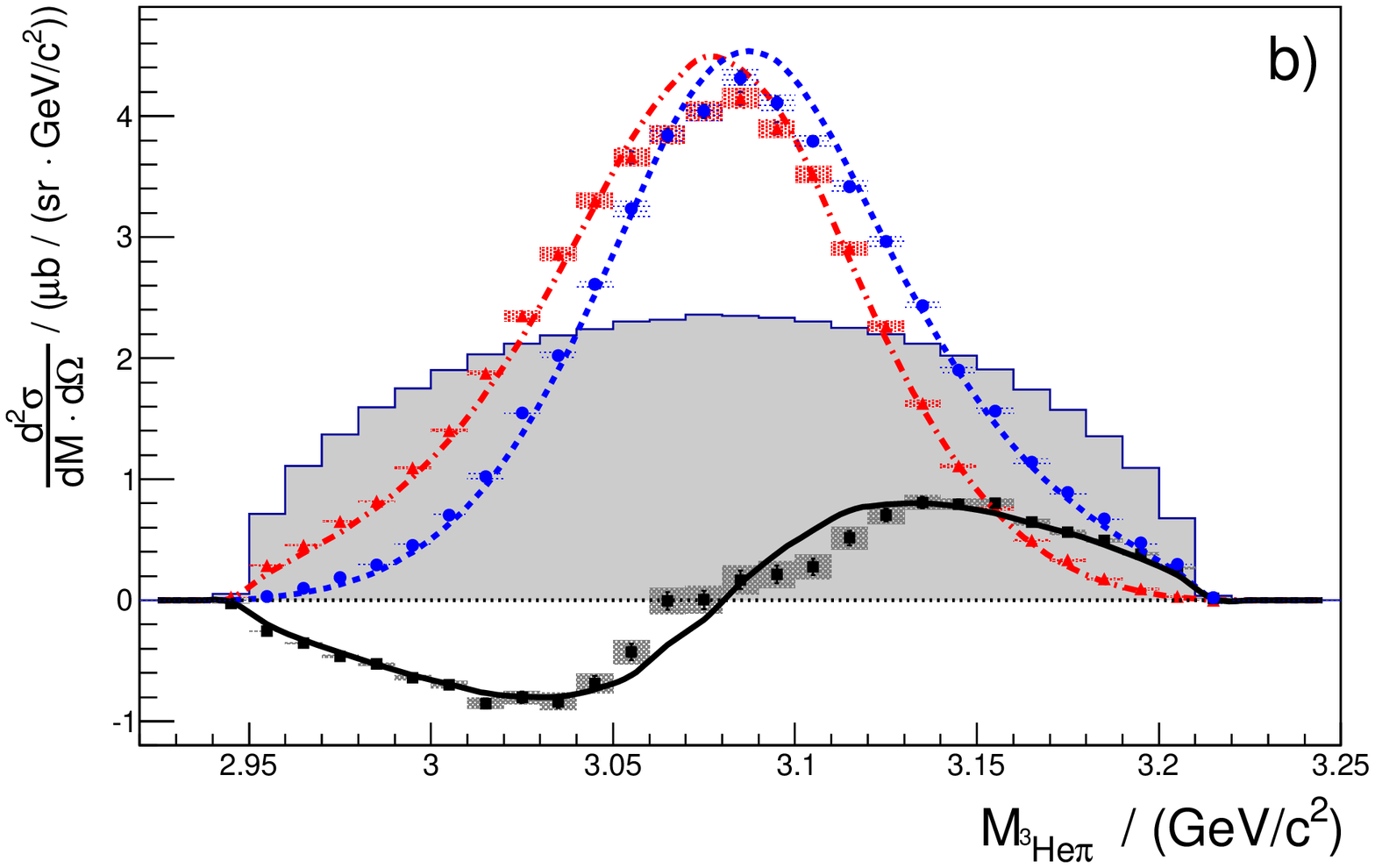}
\caption{Centre-of-mass double differential cross sections for the
\reactionb\ reaction averaged over $143^{\circ} < \vartheta^{CMS}_{^3\rm{He}}
< 173^{\circ}$ in terms of (a) $M_{\pi^+\pi^-}$ and (b)
$M_{^3\rm{He}\,\pi^+}$ (blue circles) and $M_{^3\rm{He}\,\pi^-}$ (red
triangles). The differences between the two $M_{^3\rm{He}\,\pi}$
distributions are plotted as black squares. Horizontal error bars show the
bin width, vertical ones the statistical uncertainty. The shaded rectangles
around the points reflect systematic uncertainties in the acceptance correction.
There is in addition an overall normalisation uncertainty of 6\%. The curves correspond
to Eq.~(\ref{acccorrformula1}) and the shaded areas are phase-space
distributions normalised to the integrated cross section.}	
\label{invariantmasses}
\end{figure}

Double differential cross sections for the \reactionb\ reaction are shown in
Fig.~\ref{invariantmasses} as functions of the three possible invariant mass
combinations, where the acceptance corrections have been made on the basis of
Eq.~(\ref{acccorrformula1}). In addition we also show the difference between
the $M_{^3\rm{He}\,\pi^+}$ and $M_{^3\rm{He}\,\pi^-}$ distributions.

The curves corresponding to Eq.~(\ref{acccorrformula1}) in
Fig.~\ref{invariantmasses} describe well the two $M_{^3\rm{He}\,\pi}$ spectra
and their difference but the ABC peak in the $\pi^+\pi^-$ mass distribution
is not quite sharp enough, though this could be adjusted through the
introduction of a modest $\pi^+\pi^-$ form factor.

In the approach proposed here, the isospin factors of 3 and 1 in
Eq.~(\ref{acccorrformula1}) ensure that the $\Delta(1232)$ plays a more
important role in the $M_{^3\rm{He}\,\pi^+}$ distribution than in that of the
$M_{^3\rm{He}\,\pi^-}$, with the latter being mainly a kinematic reflection
of the $\Delta^{++}(1232)$ in the available phase space. If this is true, it
would mean that a very different picture would emerge if the experiment were
repeated at significantly lower or higher energy, where the kinematic
reflections would be very different.

\begin{figure}[h]
\includegraphics[width=1.0\linewidth]{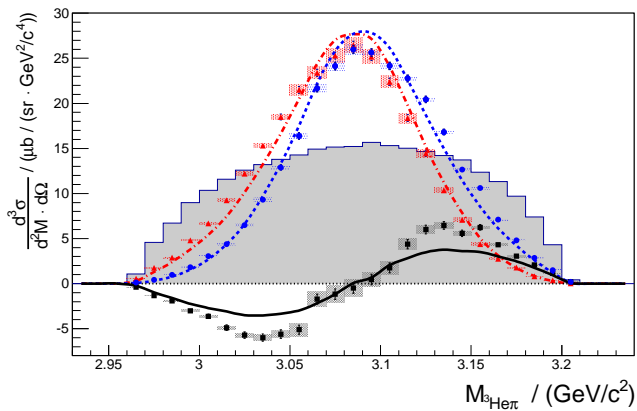}
\caption{$M_{^3\rm{He}\,\pi}$ cross section spectra for the
\reactionb\ reaction as in Fig.~\ref{invariantmasses} but only
for those events where $M_{\pi^+\pi^-} < 340$~MeV/$c^2$.}	
\label{invariantmasses_cut}
\end{figure}

However, it is clear from Fig.~(\ref{invariantmasses_cut}) that
Eq.~(\ref{acccorrformula1}) does not describe well the whole of the charge
dependence of the spectra. When events only in the ABC peak, $M_{\pi^+\pi^-}
< 340$~MeV/$c^2$, are retained, the charge difference becomes even more
pronounced than that predicted by the ansatz. The model must therefore
underestimate the strength of $I_{\pi\pi}=1$ production at low
$M_{\pi^+\pi^-}$.

%
%%%%%%%%%%%%%%%%%%%%%%%%%%%%%%%%%%%%%%%%%%%%%%%%%%%%%%%%%%%%%%%%%%%%%%%%%%%%%%
%
\section{Discussion and Conclusions}
\label{Conclusions} \setcounter{equation}{0}

Using the COSY-ANKE spectrometer, we have measured the \reactionb\ reaction
over a significant angular range at an excess energy of 265 MeV. About 81,500
fully reconstructed events were obtained with negligible background. The
results are broadly consistent with those found at CELSIUS by the WASA
collaboration~\cite{BAS2006}, though their larger geometric acceptance must
be weighed against the higher resolution and greater statistics achieved in
the current experiment.

Although ANKE has limited acceptance, the three different approaches that
have been used to compensate for this resulted in distributions that were
remarkably similar, with differences on the 5\% level. Reliable differential
cross sections could therefore be extracted for $143^{\circ} <
\vartheta_{^{3\!}{\rm He}}^{CM} < 173^{\circ}$. Over this range the
differential cross section is consistent with that found at
CELSIUS~\cite{BAS2006}.

Apart from the ABC structure, which was already well known from the inclusive
measurements~\cite{ABA1963,BAN1973}, the most striking feature of the data is
the peaking in the $M_{^3\rm{He}\,\pi^{\pm}}$ distributions. Their
difference, which was deduced with high precision, shows that there must be
some $I_{\pi\pi}=1$ $\pi^+\pi^-$ production that interferes with the dominant
$I_{\pi\pi}=0$ of the ABC enhancement. We have tried to explain this effect
in terms of a $N^*(1440) \to \Delta(1232) \to N$ decay chain and, although
this can describe the total results, it is seen from
Fig.~\ref{invariantmasses_cut} that this approach does not give sufficient
isovector strength at low $M_{\pi^+\pi^-}$. The $\Delta(1232)$ propagator
model of Eq.~(\ref{acccorrformula1}) also has the problem that it predicts no
$I_{\pi\pi}=1$ production at maximum $M_{\pi^+\pi^-}$.

Turning now to earlier experimental data, the $I_{\pi\pi}=1$ production rate
could be studied by comparing the cross sections for final $\pi^+\pi^-$ and
$\pi^0\pi^0$ states in the CELSIUS data~\cite{BAS2006}. There are some
uncertainties in the relative normalisation and, in an attempt to avoid this
problem, the data can be scaled slightly so that there is no $I_{\pi\pi}=1$
production in the phase-space-corrected data at very low $M_{\pi\pi}$.
However this then suggests little  $I_{\pi\pi}=1$ production below about
450~MeV/$c^2$ and this is certainly insufficient to explain our
$M_{^3\rm{He}\,\pi^{\pm}}$ difference, which require an $I_{\pi\pi}=1$
amplitude for $M_{\pi^+\pi^-} < 340$~MeV/$c^2$. A very reliable relative
normalisation for $\pi^0\pi^0$ and $\pi^+\pi^-$ detection is therefore a
prerequisite if one wants to use this approach to deduce the
$I_{\pi\pi}=1/I_{\pi\pi}=0$ production ratio.

If information is sought instead from the original inclusive $pd
\to{}^3\textrm{H}\,X^+/^3\textrm{He}\,X^0$ measurements~\cite{ABA1963}, there
is the difficulty of identifying the background. If this is taken to be
constant at the level found at around 270~MeV/$c^2$, then the
$I_{\pi\pi}=1/I_{\pi\pi}=0$ production ratio is over 10\% by 340~MeV/$c^2$.
Similar indications are found in the Saclay data at a deuteron momentum of
$\approx 3.4$~GeV/$c$~\cite{BAN1973}. This might be sufficient to explain the
observed $M_{^3\rm{He}\,\pi^{\pm}}$ difference, but this conclusion does
depend upon the background assumptions.

In summary, by measuring carefully the differences in the
$M_{^3\rm{He}\,\pi^+}$ and $M_{^3\rm{He}\,\pi^-}$ invariant mass spectra
produced in the \reactionb\ reaction, we have shown that there must be some
$I_{\pi\pi}=1$ production, even in the ABC region of $\pi^+\pi^-$ masses.
Similar effects cannot exist for $dd \rightarrow{}^4 \textrm{He}\,
\pi^+\pi^-$ \cite{KEL2009} because only the $I_{\pi\pi}=0$ channel is there
allowed. In the other case where the ABC effect is clearly seen, $np
\rightarrow{} \textrm{d}\, \pi^+\pi^-$, the $\pi^+d$ and $\pi^-d$ systems are
both purely $I=1$ so that little difference is to be expected in the
$M_{\pi^+d}/M_{\pi^-d}$ mass distributions, and none is observed in the WASA
data~\cite{BAS2014}.

The simple isobar production model suggested to describe our \reactionb\ data
underestimate the $I_{\pi\pi}=1$ strength at low $M_{\pi\pi}$. The effects
seem much larger than anything resulting from isospin violation and call for
a renewed effort to clarify the production mechanism in this reaction.
Further exclusive measurements at much higher or lower energy could be
very illuminating in this respect.\\

The authors wish to express their thanks to the COSY machine
crew for producing such good experimental conditions and also
to the other members of the ANKE collaboration for diverse help
in the experiment. Very generous discussions with H.~Clement
and M.~Bashkanov are also gratefully acknowledged. This work was
supported in part by the JCHP FEE.

%
%%%%%%%%%%%%%%%%%%%%%%%%%%%%%%%%%%%%%%%%%%%%%%%%%%%%%%%%%%%
%

\end{document}